\documentstyle[12pt]{article}

\newcommand {\vs}[1]  { \vspace*{#1 cm} }

\newcounter{eq}
\newcounter{sc}

\newcommand {\AP}   {Ann. of Phys.}

\newcommand {\IJMP}  {Int. J. Mod. Phys.}
\newcommand {\JHEP}   {JHEP}
\newcommand {\MPL}  {Mod. Phys. Lett.}
\newcommand {\NP}   {Nucl. Phys.}

\newcommand {\PR}   {Phys. Rev.}
\newcommand {\PRL}   {Phys. Rev. Lett.}

\def\overleftrightarrow#1{\vbox{\ialign{##\crcr
 $\leftrightarrow$\crcr\noalign{\kern-1pt\nointerlineskip}
 $\hfil\displaystyle{#1}\hfil$\crcr}}}

\newlength{\minitwocolumn}
\begin{document}

\begin{flushright}
DPUR/TH/14\\
February, 2009\\
\end{flushright}
\vspace{30pt}
\pagestyle{empty}
\baselineskip15pt

\begin{center}
{\large\bf On Unitarity of Massive Gravity in Three Dimensions
 \vskip 1mm
}

\vspace{20mm}

Masashi Nakasone
and Ichiro Oda
          \footnote{
           E-mail address:\ ioda@phys.u-ryukyu.ac.jp
                  }

\vspace{10mm}
          Department of Physics, Faculty of Science, University of the 
           Ryukyus,\\
           Nishihara, Okinawa 903-0213, JAPAN \\

\end{center}


\vspace{20mm}
\begin{abstract}
We examine a unitarity of a particular higher-derivative extension of general 
relativity in three space-time dimensions, which has been recently shown to be 
equivalent to the Pauli-Fierz massive gravity at the linearized approximation level,
and explore a possibility of generalizing the model to higher space-time dimensions.
We find that the model in three dimensions is indeed unitary in the tree-level, but
the corresponding model in higher dimensions is not so due to the appearance of 
non-unitary massless spin-2 modes. 
\vspace{15mm}

\end{abstract}

\newpage
\pagestyle{plain}
\pagenumbering{arabic}

\rm

In recent years, there has been a revival of interests of massive gravity 
models from various physical viewpoints.
{}For instance, some people conjecture that the massless graviton might
acquire mass via spontaneous symmetry breakdown of general coordinate 
reparametrization invariance, whose dynamical mechanism is sometimes called 
"gravitational Higgs mechanism" \cite{Percacci, Kaku1, Porrati, Kirsch, 't Hooft, 
Kaku2, Oda, Maeno1, Maeno2}.
This expectation naturally stems from brane world scenario where the presence 
of a brane breaks some of diffeomorphisms in the directions perpendicular to 
the brane spontaneously \cite{Kaku1, Porrati}. 
Moreover, this study is also related to the recent development of
string theory approach to quantum chromodynamics (QCD) \cite{'t Hooft}
since if we wish to apply a bosonic string theory to QCD, 
massless fields such as tachyonic scalar and spin 2 graviton in string theory, must 
become massive or be removed somehow because such the fields do not exist in QCD.

The other interest of massive gravity is relevant to the problem of counting 
the microscopic physical degrees of freedom existing in black holes through 
a holographic two-dimensional dual theory where the well-known topological massive
gravity with the Chern-Simons term \cite{Deser} plays an important role \cite{Strominger}. 

It is well-known that there is a unique way to add mass term
to general relativity in a Lorentz-covariant manner without worrying the emergence of 
a non-unitary $\it{ghost}$ in any space-time dimension whose theory is called 
the Pauli-Fierz massive gravity \cite{Fierz}. However, there is at least one serious 
disadvantage in the Pauli-Fierz massive gravity. Namely, the massive gravity
theory defined by Pauli and Fierz only makes sense as a free and linearized
theory since the diffeomorphism-invariant mass term cannot be introduced into
general relativity owing to an obvious identity $g^{\mu\nu} g_{\mu\nu} = \delta^\mu_\mu$
so it seems to be difficult to construct a sensible interacting theory for the massive 
graviton.

One resolution for overcoming this difficulty is to introduce some matter fields in general
relativity and then trigger the above-mentioned gravitational Higgs mechanism. 
However, recently, there has been an alternative progress for getting a sensible interacting
massive gravity theory in three space-time dimensions without introducing matter fields such as
scalar fields \cite{Bergshoeff}. This model has been shown to be equivalent to the Pauli-Fierz 
massive gravity at the linearized approximation level. A key idea in this model is 
that one takes into consideration higher-derivative curvature terms in the Einstein-Hilbert 
action with the $\it{wrong}$ sign in such a way that the trace part of the stress-energy tensor 
associated with those higher-derivative terms is proportional to the original 
higher-derivative Lagrangian. 

The main aim of this paper is not olny explore a possibility of generalizing this 
three-dimensional model to higher space-time dimensions but also to examine the unitarity of 
this particular higher-derivative extension of general relativity in three space-time 
dimensions by Bergshoeff et al \cite{Bergshoeff}.

Since we wish to explore a possibility of generalization of three-dimensional massive gravity
model to higher dimensions, we will start with a typical higher-derivative gravity model \cite{Stelle} 
without cosmological constant up to fourth-order in derivative in a general $D$ space-time 
dimensions \footnote{The space-time indices $\mu, \nu, \cdots$ run over $0, 1, 
\cdots, D-1$. We take the metric signature $(-, +, \cdots, +)$ and follow the notation 
and conventions of the textbook of MTW \cite{MTW}.}:  
\begin{eqnarray}
S = \int d^D x \sqrt{- g} [ - \frac{1}{\kappa^2} R  + \alpha R^2
+ \beta R_{\mu\nu} R^{\mu\nu}
+ \gamma( R_{\mu\nu\rho\sigma} R^{\mu\nu\rho\sigma} 
- 4 R_{\mu\nu} R^{\mu\nu}
+ R^2 ) ],
\label{Action}
\end{eqnarray}
where $\kappa^2 \equiv 16 \pi G_D$ ($G_D$ is the $D$-dimensional Newton's constant), 
$\alpha$, $\beta$ and $\gamma$ are constants. One important remark
is that the Einstein-Hilbert action, which is the first term having $\kappa^2$, 
has the $\it{wrong}$ sign. This is a characteristic feature in the present formalism. 
The last term proportional to $\gamma$ is nothing but the Gauss-Bonnet term, 
which is a surface term in four space-time dimensions. Einstein's equations are then 
given by
\begin{eqnarray}
- \frac{1}{\kappa^2} G_{\mu\nu} + K_{\mu\nu} = 0,
\label{Einstein's eq}
\end{eqnarray}
where $G_{\mu\nu}$ is the conventional Einstein's tensor defined as
$G_{\mu\nu} = R_{\mu\nu} - \frac{1}{2} g_{\mu\nu} R$ and the tensor
$K_{\mu\nu}$ is defined as
\begin{eqnarray}
K_{\mu\nu} &=& ( 2 \alpha + \beta ) ( g_{\mu\nu} \nabla^2 
- \nabla_\mu \nabla_\nu ) R + \beta \nabla^2 G_{\mu\nu} \nonumber\\
&+& 2 \alpha R ( R_{\mu\nu} - \frac{1}{4} g_{\mu\nu} R )
+ 2 \beta ( R_{\mu\rho\nu\sigma} - \frac{1}{4} g_{\mu\nu} R_{\rho\sigma} ) 
R^{\rho\sigma} 
+ 2 \gamma [ R R_{\mu\nu} - 2 R_{\mu\rho\nu\sigma} R^{\rho\sigma}
\nonumber\\
&+& R_{\mu\rho\sigma\tau} R_\nu \ ^{\rho\sigma\tau} - 2 R_{\mu\rho} R_\nu \ ^\rho
- \frac{1}{4} g_{\mu\nu} ( R_{\rho\sigma\tau\lambda}^2 - 4 R_{\rho\sigma}^2
+ R^2 ) ],
\label{K}
\end{eqnarray}
where $\nabla_\mu$ is the usual covariant derivative and $\nabla^2
\equiv g^{\mu\nu} \nabla_\mu \nabla_\nu$.

In the construction of a new type of massive gravity theory \cite{Bergshoeff}, 
the tensor $K_{\mu\nu}$ plays a critical role and must satisfy the following 
condition:
\begin{itemize}
\item Its trace $K \equiv g^{\mu\nu} K_{\mu\nu}$ is proportional to 
the original higher-derivative Lagrangian.
\end{itemize}
In particular, this condition ensures that the scalar curvature can be set to
zero in the trace part of the linearized Einstein's equations.

Taking trace of $K_{\mu\nu}$ gives rise to
\begin{eqnarray}
K &=& [ ( 2 \alpha + \beta ) ( D - 1) 
+ \beta ( 1 - \frac{D}{2} ) ] \nabla^2 R \nonumber\\
&+& 2 ( 1 - \frac{D}{4} ) [ \gamma R_{\mu\nu\rho\sigma}^2
+ ( \beta - 4 \gamma ) R_{\mu\nu}^2 + ( \alpha + \gamma ) R^2 ].
\label{trace K}
\end{eqnarray}
{}First, from this condition, the $\nabla^2 R$ term must vanish so that we have a relation 
between the constants $\alpha$ and $\beta$
\begin{eqnarray}
\alpha = - \frac{D}{4 (D-1)} \beta.
\label{Condition 1}
\end{eqnarray}
Then, the condition also requires three kinds of independent $R^2$ terms to satisfy
\begin{eqnarray}
(\alpha + \gamma) [ 1 - 2 c ( 1 - \frac{D}{4} ) ] &=& 0, \nonumber\\
(\beta - 4 \gamma)[ 1 - 2 c ( 1 - \frac{D}{4} ) ] &=& 0, \nonumber\\
\gamma [ 1 - 2 c ( 1 - \frac{D}{4} ) ] &=& 0,
\label{Condition 2}
\end{eqnarray}
where $c$ denotes a proportional constant.
Of course, in three and four dimensions, three $R^2$ terms are not
completely independent, so precisely speaking, the equations (\ref{Condition 2})
are valid for $D > 4$. The cases of $D = 3, \ 4$ are separately considered later.
It is obvious that all the equations in (\ref{Condition 2}) are
satisfied when 
\begin{eqnarray}
c = \frac{2}{4 - D}.
\label{C}
\end{eqnarray}
If the equation (\ref{C}) were not true, we would have a trivial
solution $\alpha = \beta = \gamma =0$, so we shall confine ourselves to
the solution (\ref{C}) in what follows. Consequently, the trace part of
the Einstein's equations (\ref{Einstein's eq}) gives us
\begin{eqnarray}
\frac{1}{\kappa^2} ( 1 - \frac{D}{2} ) R = K.
\label{trace Einstein}
\end{eqnarray}

As the next step, we shall linearize the Einstein's equations
around a Minkowski flat space-time as usual by writing out
$g_{\mu\nu} = \eta_{\mu\nu} + h_{\mu\nu}$. Eq. (\ref{trace Einstein})
together with the fact that $K$ does not involve the linear term
in $h_{\mu\nu}$ from the equation (\ref{Condition 1}) produces
\begin{eqnarray}
R^{lin} = 0,
\label{Linear Einstein1}
\end{eqnarray}
and with the help of this equation, Eq. (\ref{Einstein's eq}) yields
\begin{eqnarray}
( \Box - \frac{1}{\beta \kappa^2} ) G_{\mu\nu}^{lin} = 0,
\label{Linear Einstein2}
\end{eqnarray}
where $R^{lin}$ and $G_{\mu\nu}^{lin}$ denote the linearized scalar 
curvature and Einstein's tensor, respectively. Moreover, we have defined
$\Box \equiv \eta^{\mu\nu} \partial_\mu \partial_\nu$. Let us note that
the positivity of mass of the graviton requires us to take $\beta > 0$.

Now we are ready to show that with an appropriate choice of the constants 
$\alpha, \beta$ and $\gamma$, the action (\ref{Action}) becomes equivalent to the Pauli-Fierz 
massive gravity at the quadratic level. To do that, let us begin with an action \cite{Bergshoeff}
\begin{eqnarray}
S_f = - \frac{1}{\kappa^2} \int d^D x \sqrt{- g} [ R  + f^{\mu\nu} G_{\mu\nu}
+ \frac{m^2}{4} ( f^{\mu\nu} f_{\mu\nu} - f^2 ) ],
\label{f-action}
\end{eqnarray}
where $f_{\mu\nu}$ is some symmetric tensor field with trace $f = g^{\mu\nu} f_{\mu\nu}$ 
and $m^2$ is a constant. Integrating out the auxiliary field $f_{\mu\nu}$, 
this action is reduced to the form
\begin{eqnarray}
S_f = - \frac{1}{\kappa^2} \int d^D x \sqrt{- g}  [ R  - \frac{1}{m^2} R_{\mu\nu}^2
+ \frac{1}{m^2} \frac{D}{4(D-1)} R^2 ].
\label{f-action2}
\end{eqnarray}
Note that in order to make this action (\ref{f-action2}) coincide with 
the original action (\ref{Action}), we have to impose constraints on 
the coefficients in the action (\ref{Action})
\begin{eqnarray}
\alpha &=& - \frac{D}{4(D-1)} \frac{1}{\kappa^2 m^2}, \nonumber\\
\beta &=& \frac{1}{\kappa^2 m^2}, \nonumber\\
\gamma &=& 0.
\label{gamma}
\end{eqnarray}
It is of interest to notice that not only the first and second contraints naturally
lead to the previous relation (\ref{Condition 1}), but also the second constraint
is consistent with the mass positivity $\beta > 0$, which was mentioned below
Eq. (\ref{Linear Einstein2}).

Next, let us expand the metric around a flat Minkowski background 
$\eta_{\mu\nu}$ and keep only quadratic fluctuations in the action (\ref{f-action})
\begin{eqnarray}
S_f = \frac{1}{\kappa^2} \int d^D x [ ( f^{\mu\nu} - \frac{1}{2} h^{\mu\nu} )
{\cal{O}}_{\mu\nu, \rho\sigma} h^{\rho\sigma} 
- \frac{m^2}{4} ( f^{\mu\nu} f_{\mu\nu} - f^2 ) ],
\label{f-action3}
\end{eqnarray}
where the operator ${\cal{O}}_{\mu\nu, \rho\sigma}$ can be expressed
in terms of the spin projection operators
\begin{eqnarray}
{\cal{O}}_{\mu\nu, \rho\sigma} = \Box [ \frac{1}{2} P^{(2)} 
- \frac{D-2}{2} P^{(0, s)} ]_{\mu\nu, \rho\sigma},
\label{O}
\end{eqnarray}
where $P^{(2)}$ and $P^{(0, s)}$ are the spin-2 and spin-0 projection
operators. Concretely, in the $D$-dimensions they take the form
\begin{eqnarray}
P^{(2)}_{\mu\nu, \rho\sigma} &=& \frac{1}{2} ( \theta_{\mu\rho} \theta_{\nu\sigma}
+ \theta_{\mu\sigma} \theta_{\nu\rho} ) - \frac{1}{D-1} \theta_{\mu\nu} \theta_{\rho\sigma},
\nonumber\\  
P^{(0, s)}_{\mu\nu, \rho\sigma} &=& \frac{1}{D-1} \theta_{\mu\nu} \theta_{\rho\sigma},
\label{Spin projectors}
\end{eqnarray}
where the transverse operator $\theta_{\mu\nu}$ and the longitudinal operator
$\omega_{\mu\nu}$ are defined as
\begin{eqnarray}
\theta_{\mu\nu} &=& \eta_{\mu\nu} - \frac{1}{\Box} \partial_\mu \partial_\nu 
= \eta_{\mu\nu} - \omega_{\mu\nu}, \nonumber\\  
\omega_{\mu\nu} &=& \frac{1}{\Box} \partial_\mu \partial_\nu.
\label{theta}
\end{eqnarray}
It is worthwhile to stress that the structure of the operator 
${\cal{O}}_{\mu\nu, \rho\sigma}$ is the same in any space-time dimension.

Here we wish to perform the path integration over $h_{\mu\nu}$ in the action 
(\ref{f-action3}). To do that, it is convenient to think of partition function
\begin{eqnarray}
Z = \int {\cal{D}} h_{\mu\nu} {\cal{D}} f_{\mu\nu} e^{ i S_f }.
\label{Z}
\end{eqnarray}
One can rewrite the action (\ref{f-action3}) as
\begin{eqnarray}
S_f = \frac{1}{\kappa^2} \int d^D x [ - \frac{1}{2} ( h - f )^{\mu\nu} 
{\cal{O}}_{\mu\nu, \rho\sigma} ( h - f )^{\rho\sigma}
+ \frac{1}{2} f^{\mu\nu} {\cal{O}}_{\mu\nu, \rho\sigma} f^{\rho\sigma} 
- \frac{m^2}{4} ( f^{\mu\nu} f_{\mu\nu} - f^2 ) ].
\label{f-action4}
\end{eqnarray}
Changing the variables from $h_{\mu\nu}$ to $k_{\mu\nu} \equiv h_{\mu\nu} - f_{\mu\nu}$,
the partition function (\ref{Z}) reads
\begin{eqnarray}
Z = \int {\cal{D}} k_{\mu\nu} {\cal{D}} f_{\mu\nu} e^{ i S^\prime_f },
\label{Z2}
\end{eqnarray}
where $S'_f$ is defined by
\begin{eqnarray}
S^\prime_f \equiv \frac{1}{\kappa^2} \int d^D x [ - \frac{1}{2} k^{\mu\nu} 
{\cal{O}}_{\mu\nu, \rho\sigma} k^{\rho\sigma}
+ \frac{1}{2} f^{\mu\nu} {\cal{O}}_{\mu\nu, \rho\sigma} f^{\rho\sigma} 
- \frac{m^2}{4} ( f^{\mu\nu} f_{\mu\nu} - f^2 ) ].
\label{f'-action}
\end{eqnarray}

In attempting to perform the path integration over $k^{\mu\nu}$, we find it 
impossible to do so since there is no inverse matrix of ${\cal{O}}_{\mu\nu, \rho\sigma}$.
That is, because of the gauge invariance, the linearized diffeomorphisms, 
in the action (\ref{f'-action}), the operator ${\cal{O}}_{\mu\nu, \rho\sigma}$
has zero eigenvalues so that its inverse matrix, which is in essence the propagator
of the massless graviton, does not generally exist. This is also clear from 
the observation that we need more spin projection operators 
$P^{(1)}, P^{(0, w)}, P^{(0, sw)}$ and $P^{(0, ws)}$ in addition to 
$P^{(2)}$ and $P^{(0, s)}$ in order to form a complete set of 
the spin projection operators in the space of second rank symmetric tensors. 
Thus, in order to make the operator ${\cal{O}}_{\mu\nu, \rho\sigma}$ invertible,
we fix the gauge transformations by the De Donder's gauge-fixing conditions.
Then, the gauge-fixed action of (\ref{f'-action}) is of form
\begin{eqnarray}
\hat{S}_f &\equiv& \frac{1}{\kappa^2} \int d^D x [ - \frac{1}{2} k^{\mu\nu} 
{\cal{O}}_{\mu\nu, \rho\sigma} k^{\rho\sigma}
+ \frac{1}{2 \alpha} ( \partial_\nu k_\mu \ ^\nu - \frac{1}{2} \partial_\mu k )^2
\nonumber\\
&+& \frac{1}{2} f^{\mu\nu} {\cal{O}}_{\mu\nu, \rho\sigma} f^{\rho\sigma} 
- \frac{m^2}{4} ( f^{\mu\nu} f_{\mu\nu} - f^2 ) ]
\nonumber\\
&=& \frac{1}{\kappa^2} \int d^D x [ - \frac{1}{2} k^{\mu\nu} 
\hat{{\cal{O}}}_{\mu\nu, \rho\sigma} k^{\rho\sigma}
+ \frac{1}{2} f^{\mu\nu} {\cal{O}}_{\mu\nu, \rho\sigma} f^{\rho\sigma} 
- \frac{m^2}{4} ( f^{\mu\nu} f_{\mu\nu} - f^2 ) ],
\label{f'-action2}
\end{eqnarray}
where $\alpha$ is a gauge parameter and the new operator 
$\hat{{\cal{O}}}_{\mu\nu, \rho\sigma}$ is defined through a complete set of 
the spin projection operators 
\begin{eqnarray}
\hat{{\cal{O}}}_{\mu\nu, \rho\sigma} &=& \Box [ \frac{1}{2} P^{(2)}
+  \frac{1}{2 \alpha} P^{(1)} + \frac{-2(D-2)\alpha + D - 1}{4 \alpha} P^{(0, s)} 
+ \frac{1}{4 \alpha} P^{(0, w)} 
\nonumber\\
&-& \frac{\sqrt{D-1}}{4 \alpha} P^{(0, sw)} 
- \frac{\sqrt{D-1}}{4 \alpha} P^{(0, ws)} ]_{\mu\nu, \rho\sigma},
\label{hat O}
\end{eqnarray}
where $P^{(1)}, P^{(0, w)}, P^{(0, sw)}$ and $P^{(0, ws)}$ are defined as
\begin{eqnarray}
P^{(1)}_{\mu\nu, \rho\sigma} &=& \frac{1}{2} ( \theta_{\mu\rho} \omega_{\nu\sigma}
+ \theta_{\mu\sigma} \omega_{\nu\rho} + \theta_{\nu\rho} \omega_{\mu\sigma}
+ \theta_{\nu\sigma} \omega_{\mu\rho} ),
\nonumber\\  
P^{(0, w)}_{\mu\nu, \rho\sigma} &=& \omega_{\mu\nu} \omega_{\rho\sigma},
\nonumber\\  
P^{(0, sw)}_{\mu\nu, \rho\sigma} &=& \frac{1}{\sqrt{D-1}} \theta_{\mu\nu} \omega_{\rho\sigma},
\nonumber\\  
P^{(0, ws)}_{\mu\nu, \rho\sigma} &=& \frac{1}{\sqrt{D-1}} \omega_{\mu\nu} \theta_{\rho\sigma}.
\label{Spin projectors2}
\end{eqnarray}
Note that all the spin projection operators $\{ P^{(2)}, P^{(1)}, P^{(0, s)}, P^{(0, w)},
P^{(0, sw)}, P^{(0, ws)} \}$ satisfy the orthogonality relations
\begin{eqnarray}
P_{\mu\nu, \rho\sigma}^{(i, a)} P_{\rho\sigma, \lambda\tau}^{(j, b)}
&=& \delta^{ij} \delta^{ab} P_{\mu\nu, \lambda\tau}^{(i, a)},
\nonumber\\  
P_{\mu\nu, \rho\sigma}^{(i, ab)} P_{\rho\sigma, \lambda\tau}^{(j, cd)}
&=& \delta^{ij} \delta^{bc} P_{\mu\nu, \lambda\tau}^{(i, a)},
\nonumber\\  
P_{\mu\nu, \rho\sigma}^{(i, a)} P_{\rho\sigma, \lambda\tau}^{(j, bc)}
&=& \delta^{ij} \delta^{ab} P_{\mu\nu, \lambda\tau}^{(i, ac)},
\nonumber\\  
P_{\mu\nu, \rho\sigma}^{(i, ab)} P_{\rho\sigma, \lambda\tau}^{(j, c)}
&=& \delta^{ij} \delta^{bc} P_{\mu\nu, \lambda\tau}^{(i, ac)},
\label{Orthogonality}
\end{eqnarray}
with $i, j = 0, 1, 2$ and $a, b, c, d = s, w$ and the tensorial relation 
\begin{eqnarray}
[ P^{(2)} + P^{(1)} + P^{(0, s)} + P^{(0, w)} ]_{\mu\nu, \rho\sigma} = 
\frac{1}{2} ( \eta_{\mu\rho}\eta_{\nu\sigma} + \eta_{\mu\sigma}\eta_{\nu\rho} ).
\label{T relation}
\end{eqnarray}
Using these relations, it is straightforward to derive the inverse of 
the matrix $\hat{{\cal{O}}}_{\mu\nu, \rho\sigma}$
\begin{eqnarray}
\hat{{\cal{O}}}_{\mu\nu, \rho\sigma}^{-1} &=& \frac{1}{\Box}
[ 2 P^{(2)} +  2 \alpha P^{(1)} - \frac{2}{D-2} P^{(0, s)} 
- \frac{2 \{ -2(D-2)\alpha + D - 1 \}}{D-2} P^{(0, w)} 
\nonumber\\
&-& \frac{2\sqrt{D-1}}{D-2} P^{(0, sw)} 
- \frac{2\sqrt{D-1}}{D-2} P^{(0, ws)} ]_{\mu\nu, \rho\sigma}.
\label{Inverse O}
\end{eqnarray}
Hence, we can now perform the path integration over $k_{\mu\nu}$ without
hesitation 
\begin{eqnarray}
Z = \int {\cal{D}} f_{\mu\nu} e^{ i S_{PF} },
\label{Z3}
\end{eqnarray}
where $S_{PF}$ is the Pauli-Fierz massive gravity action with the $\it{correct}$ 
sign \cite{Fierz}:
\begin{eqnarray}
S_{PF} = \frac{1}{\kappa^2} \int d^D x [ \frac{1}{2} f^{\mu\nu}
{\cal{O}}_{\mu\nu, \rho\sigma} f^{\rho\sigma} 
- \frac{m^2}{4} ( f^{\mu\nu} f_{\mu\nu} - f^2 ) ].
\label{final action}
\end{eqnarray}

Let us consider deliberately what we have done above. It seems that  
the action (\ref{f-action2}), or equivalently the action (\ref{f-action}),
is equivalent to the Pauli-Fierz massive gravity action (\ref{final action}) 
at least at the linearized level since, as seen in Eq. (\ref{f'-action}), 
the tensor field $k_{\mu\nu}$ does not interact with the other tensor one 
$f_{\mu\nu}$ at all so that we can integrate $k_{\mu\nu}$ away after the gauge-fixing. 
However, in the higher-order approximation level, there appear interaction 
terms between $k_{\mu\nu}$ (in other words, $h_{\mu\nu}$)
and $f_{\mu\nu}$, so that it is impossible to perform the path integration
over $k_{\mu\nu}$ to arrive at the Pauli-Fierz action (\ref{final action}).
Then, we can only show that the action (\ref{f-action2}) $\it{is}$ equivalent 
to the action $S'_f$ in (\ref{f'-action}) with many of interaction terms
involving the tensor fields $k_{\mu\nu}$ and $f_{\mu\nu}$, which is essentially
an interacting theory of two symmetric tensor fields where one is a massless 
tensor field with the $\it{wrong}$ sign and the other is a massive tensor one with the 
$\it{correct}$ sign. Incidentally, let us mention the cases of three ($D=3$) and four 
($D=4$) dimensional space-time. In the three dimensional case, it is easy
to see that the present analysis naturally reduces to the work by 
Bergshoeff et al \cite{Bergshoeff}. On the other hand, in the case of four dimensions,
we find it impossible to construct the tensor $K_{\mu\nu}$ satisfying the condition.
This fact can be also seen in the presence of the pole at $D=4$ in Eq. (\ref{C}).

To give a definite answer to a question whether or not our massive gravity model is really 
physically plausible, we have to investigate the property of unitarity of the higher-derivative 
action (\ref{f-action2}) directly. Actually, we will find that the model is unitary only in three 
dimensions while in the other dimensions we have non-unitary massless spin-2 modes which come from 
the $\it{wrong}$ sign in front of the Einstein-Hilbert action. Thus, it is impossible to
generalize the three-dimensional massive gravity model by Bergshoeff et al. \cite{Bergshoeff} 
to higher space-time dimensions.

To this aim, let us notice that each term in the action (\ref{f-action2}) is expressed 
by the spin projection operators as
\begin{eqnarray}
- \sqrt{- g} R &=& - \frac{1}{4} h^{\mu\nu} [ P^{(2)} - (D-2) P^{(0, s)} ]_{\mu\nu, \rho\sigma}
\Box h^{\rho\sigma},
\nonumber\\
\xi \sqrt{- g} R_{\mu\nu} R^{\mu\nu} 
&=& \xi \frac{1}{4} h^{\mu\nu} [ P^{(2)} + D P^{(0, s)} ]_{\mu\nu, \rho\sigma}
\Box^2 h^{\rho\sigma},
\nonumber\\
\lambda \sqrt{- g} R^2 
&=& - \xi \frac{D}{4} h^{\mu\nu} P^{(0, s)}_{\mu\nu, \rho\sigma}
\Box^2 h^{\rho\sigma},
\label{Lagrangian}
\end{eqnarray}
where we have defined as $\xi \equiv \frac{1}{m^2}$ and $\lambda \equiv 
- \frac{1}{m^2} \frac{D}{4(D-1)} \equiv - \frac{D}{4(D-1)} \xi$.
A nice feature in the present theory is that with the coefficients in front of
the higher-derivative terms, the scalar ghost mode which exists in the 
spin projection operator $P^{(0, s)}$ is canceled out as can be seen 
\begin{eqnarray}
\xi \sqrt{- g} R_{\mu\nu} R^{\mu\nu} + \lambda \sqrt{- g} R^2 
= \xi \frac{1}{4} h^{\mu\nu} P^{(2)}_{\mu\nu, \rho\sigma} \Box^2 h^{\rho\sigma}.
\label{Cancellation}
\end{eqnarray}
Taking the De Donder's gauge conditions for diffeomorphisms again, the 
quadratic Lagrangian part of the action (\ref{f-action2}) (up to the overall
constant $\frac{1}{\kappa^2}$) reads
\begin{eqnarray}
{\cal{L}} = \frac{1}{2} h^{\mu\nu} {\cal{P}}_{\mu\nu, \rho\sigma} h^{\rho\sigma},
\label{Quadratic}
\end{eqnarray}
where the operator $\cal{P}$ is defined as 
\begin{eqnarray}
{\cal{P}}_{\mu\nu, \rho\sigma} &=& \Box [ \frac{1}{2} (-1 + \xi \Box) P^{(2)}
-  \frac{1}{2 \alpha} P^{(1)} + \frac{2(D-2) \alpha - (D - 1)}{4 \alpha} P^{(0, s)} 
- \frac{1}{4 \alpha} P^{(0, w)} 
\nonumber\\
&+& \frac{\sqrt{D-1}}{4 \alpha} P^{(0, sw)} 
+ \frac{\sqrt{D-1}}{4 \alpha} P^{(0, ws)} ]_{\mu\nu, \rho\sigma}.
\label{P}
\end{eqnarray}
Then, the inverse of the operator $\cal{P}$ is calculated as
\begin{eqnarray}
{\cal{P}}_{\mu\nu, \rho\sigma}^{-1} &=& \frac{1}{\Box} 
[ \frac{2}{-1 + \xi \Box} P^{(2)} -  2 \alpha P^{(1)} 
+ \frac{2}{D-2} P^{(0, s)} + \frac{2(D-1) - 4 \alpha(D-2)}{D-2} P^{(0, w)} 
\nonumber\\
&+& \frac{2 \sqrt{D-1}}{D-2} P^{(0, sw)} 
+ \frac{2 \sqrt{D-1}}{D-2} P^{(0, ws)} ]_{\mu\nu, \rho\sigma}.
\label{P-inv}
\end{eqnarray}
Using it, the propagator for $h_{\mu\nu}$ takes the form
\begin{eqnarray}
<0| T (h_{\mu\nu}(x) h_{\rho\sigma}(y)) |0>
= i {\cal{P}}_{\mu\nu, \rho\sigma}^{-1} \delta^{(D)}(x-y).
\label{Propa}
\end{eqnarray}

Now we are willing to investigate the unitarity of the theory. One of the
easiest way is to see the imaginary part of the residue of the tree-level 
amplitudes at the poles where the external sources are conserved, 
transverse stress-energy tensor. Then, the longitudinal operator $\omega_{\mu\nu}$
in the spin projector operators does not contribute, so only the projection
operators $P^{(2)}$ and $P^{(0, s)}$ survive. Thus, the amplitude $A$
takes the form in the momentum space
\begin{eqnarray}
A &=& i T^{* \mu\nu} [ \frac{2}{p^2 + \frac{1}{\xi}} P^{(2)}
- \frac{2}{p^2} ( P^{(2)} - \frac{1}{D-2} P^{(0, s)} ) ]_{\mu\nu, \rho\sigma}
T^{\rho\sigma}
\nonumber\\
&=& i [ \frac{2}{p^2 + \frac{1}{\xi}} ( |T_{\mu\nu}|^2 - \frac{1}{D-1} |T^\mu_\mu|^2 )
- \frac{2}{p^2} ( |T_{\mu\nu}|^2 - \frac{1}{D-2} |T^\mu_\mu|^2 ) ].
\label{Amplitude}
\end{eqnarray}
Since the stress-energy tensor $T_{\mu\nu}$ is now conserved and
transverse, we can expand it in terms of the polarization vector $\varepsilon_\mu^i$ 
with $i = 1, 2, \cdots, D-2$ as $T_{\mu\nu} = t_{ij} \varepsilon_\mu^i \varepsilon_\nu^j$.
Then the amplitude $A$ can be rewritten as 
\begin{eqnarray}
A = i [ \frac{2}{p^2 + \frac{1}{\xi}} ( |t_{ij}|^2 - \frac{1}{D-1} |t^i_i|^2 )
- \frac{2}{p^2} ( |t_{ij}|^2 - \frac{1}{D-2} |t^i_i|^2 ) ].
\label{Amplitude2}
\end{eqnarray}

It is now straightforward to evaluate the imaginary part of the residue of 
the amplitude at the poles. First, at the massless pole corresponding to
the massless graviton, we have
\begin{eqnarray}
Im Res (A)|_{p^2 = 0} = -2 ( |t_{ij}|^2 - \frac{1}{D-2} |t^i_i|^2 ) ].
\label{Pole1}
\end{eqnarray}
This is obviously vanishing for $D=3$ while it becomes negative for $D>4$.
This fact implies that there is no dynamical massless graviton in three
dimensions whereas the massless graviton becomes a $\it{ghost}$ for $D>4$.
On the other hand, at the massive pole corresponding to the massive graviton
\begin{eqnarray}
Im Res (A)|_{p^2 = - \frac{1}{\xi}} = 2 ( |t_{ij}|^2 - \frac{1}{D-1} |t^i_i|^2 ) ],
\label{Pole2}
\end{eqnarray}
which is positive for both $D=3$ and $D>4$. Therefore, the massive graviton
with mass '$m$' is a dynamical field with the positive norm. Accordingly,
it is worthwhile to emphasize that the gravitational theory defined by the
action (\ref{f-action2}) is free from the ghost and describes 
an interacting unitary massive gravity theory only in three space-time dimensions 
whereas it is not unitary
in more than four dimensions. This fact is also certified by the observation that
the action (\ref{f'-action}) includes the Einstein-Hilbert action with the 
$\it{wrong}$ sign so that the corresponding massless graviton mode has a negative
norm, which is non-dynamical only in three dimensions. It is remarkable that
the ghost does not show up in three dimensions even if there is a propagator
like $\frac{1}{\Box ( \Box - \frac{1}{\xi} )}$ which can be seen in Eq. (\ref{P-inv}). 

In this short article, we have clarified unitarity of a massive gravity model
in three space-time dimensions by Bergshoeff et al \cite{Bergshoeff}.
Although we can formally construct a sort of dual action (\ref{f-action}) which connects 
the higher-derivative action (\ref{f-action2}) and the Pauli-Fierz massive gravity 
action (\ref{final action}) at the quadratic level via path integration,
it has turned out that it is necessary to analyze the higher-derivative action
(\ref{f-action2}) in some detail. We have pointed out that even if it seems to be 
possible to generalize the three-dimensional theory with a particular higher-derivative
terms \cite{Bergshoeff}, to higher dimensions except in four dimensions,
only the three-dimensional theory provides a unitary theory of the massive graviton. 
This is because in three space-time dimensions the non-unitary massless graviton
mode is not dynamical while in the other higher dimensions it becomes dynamical,
thereby violating the unitarity of the theory.

\vs 1   

\end{document}